# Dpush: A scalable decentralized spam resistant unsolicited messaging protocol.

Sam Maloney
Dmail: samzu1ctt7kscitkrt5jft91gtw5c1i6
sam2015@morph.is
https://morph.is

Paper UUID: 6e0dc694-ed49-46ab-8c16-ad65828fd533

**Abstract**: Herein this paper is presented a novel invention – called Dpush – that enables truly scalable spam resistant uncensorable automatically encrypted and inherently authenticated messaging; thus restoring our ability to exert our right to private communication, and thus a step forward in restoring an uncorrupted democracy. Using a novel combination of a distributed hash table[1] (DHT) and a proof of work[2] (POW), combined in a way that can only be called a synergy, the emergent property of a scalable and spam resistant unsolicited messaging protocol elegantly emerges. Notable is that the receiver does not need to be online at the time the message is sent. This invention is already implemented and operating within the package that is called MORPHiS – which is a Sybil[3] resistant enhanced Kademlia[1] DHT implementation combined with an already functioning implementation of Dpush, as well as a polished HTTP Dmail interface to send and receive such messages today. MORPHiS is available for free (GPLv2) at the https://morph.is website.

## 1. Introduction

Decentralization of everything is necessary. Centralization of any thing has the inherent flaw that that centralized entity will then have a position of information asymmetry. Information is power, and thus it is the inherent nature of a centralized entity to leverage that power to its selfish advantage. John Dalbert-Acton pointed out the fact that power tends towards corruption, and everything secret degenerates. This is inherent in a world bound by the laws of thermodynamics and entropy. To prevent this inherent corruption of centralized entities from exerting their degenerate will upon humanity, we must deprecate all excuses for their existence. The solution is to decentralize everything.

Humans since creation had the ability to converse in private without the ability for centralized entities to spy on that communication and thus build up an information asymmetry against the general population. Such information asymmetry completely subverts democracy, and free will. Thus, private uncensorable communication was a necessary right bestowed upon us by our Creator. We were always able to whisper to each other, without third parties being able to eavesdrop. Eons later, technology appeared – for only a very short time now relative to our total existence – that offered various means of communication which extended the range that was previously possible. Every such advance however brought the disadvantage that such communication was no longer secure against eavesdropping and thus centralized entities from building up an arsenal of information asymmetry to wield against others which they then inherently use to enslave humanity for said entities own selfish degenerate interests.

Herein this paper is presented a novel invention – called Dpush – that enables truly scalable spam resistant uncensorable automatically encrypted and inherently authenticated messaging; thus restoring our ability to exert our right to private communication, and thus providing a step forward in restoring an uncorrupted democracy. Using a novel combination of a distributed hash table[1] (DHT) and a proof of work[2] (POW), combined in a way that can only be called a synergy, the emergent property of a scalable and spam resistant unsolicited messaging protocol elegantly emerges.

By unsolicited messaging, it is meant that the sender of a message does not have to have given the receiver any prior information in order for the receiver to see the message sent. The receiver need only once publish a target_key that they wish messages to be sent to. That is the key space in the DHT where they will be looking for new messages. Such look up is extremely efficient, as the network being a DHT, is a log scaling distributed database that scales log base anything desired. The base is able to be controlled by a memory tradeoff which for example provides a network of 7 billion nodes with an average hop distance of less than 3 with a memory usage of less than 256 kilobytes for the routing table. With a 16 megabyte routing table, less than 2 average hops to reach over 7 billion nodes is possible. The receiver of messages over the Dpush system only sees the messages that completed enough proof of work to get into their published target_key key space. There is no filtering of spam necessary. Due to the fact that all of the work required is done on the sender's computer, no denial of service (DOS) attack is possible on the target_key key space. Neither the network nor the receiver is burdened with having to sort through junk or fake messages. The messages that did not complete enough work on the senders computer simply do not get into the target_key key space and thus do not affect network performance in the slightest.

Just as how Bitcoin is the unique invention of combining a simple POW with the concept of a linked list of hash trees, and through such synergy of concepts the emergent property of a decentralized and useful currency solving a previously open problem emerged; Dpush is the unique invention of combining a DHT with a variation of POW that is in this case implemented in a way that is inherent to the nature of a DHT, and through such a synergy the emergent property of a spam resistant unsolicited messaging system emerges.

In a further invention enabled by Dpush that is called Dmail, which adds automatic fool-proof encryption and inherent authentication to the messaging protocol, this target_key may be published by signing it with the asymmetric key whose public component is the data hashed to determine the Dmail address. Dmail is a very simple use of Dpush and may be described in a follow up paper; however, it is so simple that it will be described later in this paper as well. The mechanism already described above allows the target_key to be rotated without the Dmail address changing. With Dpush, the receiver has control over the target_key, the difficulty of the proof of work necessary for a message to be sent to their target_key, as well as any other instructions for sending that message, including what encryption to use, if any, as well as any ephemeral keys to use for that encryption. Any meta information can also be stored in the updateable key data signed by the receivers public key.

Dpush can be used many ways to solve many related problems – wherever the root of the problem is enabling scalable spam and DOS proof unsolicited messaging. The next planned implementation of Dpush technology will be to provide the world with the distributed discussion system (DDS). This will deprecate web comment systems such as Disqus, as well as any forum system, such as Reddit. DDS has not been implemented as of yet at the time of the writing of this paper; however, it is a very simple additional layer – just like Dmail – and like Dmail, it requires no new real inventions beyond the existing invention of Dpush. Thus, it will not take much time to achieve. Many such things are already thought out, and in the time to come will be implemented and provided free to the world for the good of all mankind.

## 2. Spam Resistance

The invention described in this paper provides a mechanism for sending unsolicited messages where the sender has the inherent need to perform a proof of work inherent to the nature of a DHT in order for their message to be stored where it will be seen and thus later retrieved by a receiver. This work is completely limited to the sender's computer. Other than the trap door verification step performed by the network and the receiver, there is no cost to the receiver or network relative to what the sender must do for their message to succeed.

A study[3] of spam marketing conversion rates by teams at Berkeley and at the University of California noted that amongst 350 million real e-mail spam messages that they observed, there were only 28 conversions (a rate of well under 0.00001%). This is 12,500,000 spam e-mails having to be sent for just one single sale. This means that the work necessary for a spammer to make one sale, and thus achieve their objective, is 12.5 million times the work that a legitimate mail sender need perform. This means for example that if just 5 seconds of computer work is set as the average difficulty of the proof of works required to send messages to a list, then a spammer has to do 1.98 computer years of processing to make each single sale. With even such a low setting of difficulty, which as mentioned is easily configurable on a per address basis by each receiver, spam becomes completely economically infeasible.

Included in the message received by the receiver can be an ID in the DHT key space of an updateable key controlled by the sender. The receiver can add such an ID (updateable site) of the sender to an address book that is periodically scanned for new messages. The sender may then send additional messages with no proof of work required by simply publishing additional messages to the signed key space that is such an updateable key that they control by having the private key that signs it.

Similarly, in order to send a solicited mass newsletter, mailing-list, Etc., where the receivers are subscribed instead of unsoliciting, the amount of work necessary to send a message to n amount of such receivers is O(1), and less than that required for sending one unsolicited message. This is because no proof of work is necessary to send to subscribed receivers. The subscribers simply query the DHT for the data stored under a signed key space published by the sender.

## 3. TargetedBlock

TargetedBlock is the name of the core underlying data structure of Dpush. It is what enables the scalable inherently spam resistant unsolicited messaging ability of the higher level Dpush.

TargetedBlock is structured as follows: A header which contains a target_key, a nonce value, and a hash of its data block; and then the data block itself. This allows the proof of work to be independent of the size of the message. By using a hash tree, the message can be any size, and yet only the first block needs be checked to be verified before accepting or downloading any more data.

```
[  TargetedBlock      ]
[----- header:        ]
[nonce                ]
[target_key           ]
```

```
[block_hash         ]
[----- data:        ]
[block(data)        ]
```

The target_key is necessary to prevent mass targeting by doing the proof of work and comparing the resulting hash against a list of target_keys that one is attempting to spam. By forcing the target_key being targeted to be part of the data being hashed as the proof of work, the DHT (network) and the receiver can ignore any TargetedBlock where the proof of work was not done itself targeting a specific single target_key.

The nonce is necessary to enable the mechanism of the inherent proof of work synergized with the DHT.

In a DHT that supports having the ID that data is stored under to be a hash of the data, such a DHT need only have a very slight additional code path to support the TargetedBlock concept. The DHT instead stores the TargetedBlock under the ID that is a hash of just the TargetedBlock's header.

The attached data is then hashed and checked against the block_hash value in the header before storing the TargetedBlock as a whole. This enables the difficulty of the proof of work to be independent of the amount of data being sent and stored. A DHT would impose a reasonable bounds to the size of the attached data by the inherent block size of the DHT. Larger messages can be sent by having the TargetedBlock data a hash tree referring to additional data that is uploaded to the network by a standard as opposed to TargetedBlock (and thus POW) means.

For a sender to send an unsolicited message, all they need is the target_key and difficulty setting of the receiver. The sender then builds a TargetedBlock with their desired message data, and a valid header. They then hash the header and then they check the hash result against the target_key and difficulty. If the resulting hash is between the target_key and target_key + difficulty, then the TargetedBlock is now valid. If not, the receiver increments the nonce value in the header and hashes the header again. They repeat this step until a valid header is obtained. When the TargetedBlock header is valid, the network will now store it, and more importantly, the receiver will see the message when they query the DHT network for their target_key keyspace. The store operation is as efficient and scalable as a normal DHT put operation. Notable is that the receiver does not need to be online at the time the message is sent.

The DHT also needs the simple additional feature of being able to be queried by a range of IDs. This is simple and natural in most DHT implementations. The query, instead of being done for a full ID, is done by asking for a common prefix of the target_key masked with its difficulty bits. As long as the difficulty is high enough that the resulting range is narrower than the key space that the average node stores, than look-up success is guaranteed with the same efficiency as a normal DHT key look up.

For a receiver to find the messages that have been sent to them, all they have to do is ask the network for any data that is >= their target_key and < their target_key + difficulty. This is a range based DHT get operation. With a high enough difficulty setting (narrow enough key space), the network is able to do this query as efficiently, as accurately, and as scalable as a normal DHT get operation. If there are no messages, then the get operation returns as efficiently as a normal query that results in no data found. If there is a message found, then that message is returned to the receiver instantly. The receiver then repeats or continues the query but instead of the target_key as the ID for the network to find, they specify the ID + 1 of the returned message. They still specify the original target_key as additional data to the query so that the network only returns TargetedBlock entries that contain that exact target_key in the TargetedBlock header. When a target_key space has more than a reasonable amount of messages, making finding the farthest messages take more operations than is desired, the receiver can publish a new target_key. The receiver can still check previous target_key values for some time to ensure that no messages are lost due to senders using an outdated value of the target_key. The receiver need only connect to the network in order to retrieve the messages from the DHT; they do not need to be connected at the time the messages are sent to the network by the sender.

**4. Dpush**

      Dpush is taking the TargetedBlock concept a tiny bit further to make a higher level general purpose unsolicited messaging protocol. Dpush can work many ways, but the basis and current implementation is described as follows.

      Since the TargetedBlock concept is a synergy of a DHT with a POW in order for the Dpush property to emerge, it is only natural that the further abilities of Dpush take advantage of what any decent modern DHT implementation can provide. Dpush takes the TargetedBlock invention and combines it with the updateable key concept in order to provide a simple and powerful protocol for unsolicited messaging in a decentralized manner.

      By updateable key concept, it is meant the concept of instead of having data stored in a DHT under an ID that is a hash of the data, the data is signed and stored under the hash of the public key that signed the data. Some DHT implementations of this concept use the public key itself as the ID. However, in the DHT implementation that is MORPHiS, it was decided to use the hash of the public key – in order to abstract the asymmetric encryption technology from the DHT key space. MORPHiS will be described in a follow up paper to come later, as well as briefly in Section 6 below.

      Dpush works by having the receiver (address owner) publish an updateable key (Dpush address) which stores meta-data (Dpush site) that is used by the sender wishing to send said receiver an unsolicited message. This meta-data as implemented in MORPHiS is simply a JSON text, so as to allow for easy extensibility. This data includes the necessary target_key property. The target_key is the real ID inside the DHT's key space that the receiver queries in order to find messages sent to them. However, in order to support much more power than the receiver simply advertising a single target_key directly, Dpush uses the Dpush site concept to enable many beneficial things. One great immediate benefit is that this mechanism allows the receiver to rotate the target_key that they will be checking, and to do so without their Dpush address from ever needing to change. This allows the receiver to switch to a new target_key whenever the previous one has become cluttered with (even valid) messages and thus prevent the gradual slowing down of the scanning of that key space. Because the target_key is specified in the TargetedBlock header, the target_key need only be incremented by one least significant bit in order to create a completely new key space for all intents and purposes. This indirection also allows multiple target_key values to be listed in one Dpush address by the address owner. This allows the receiver to query multiple keys in parallel to speed up message download if they are a high volume receiver. They could also specify multiple target_key values with each a different difficulty setting, in order to enable different priority mailboxes all with the same Dpush address. The data published in the Dpush site also importantly contains the difficulty setting that the address owner has currently specified. This allows the difficulty to be set on a per address or even per target_key basis, and changed at any time at the whim of the address owner. The difficulty setting as currently implemented in MORPHiS Dpush is that it specifies how many leading bits of the target_key must be matched by the

resulting hash of the TargetedBlock header in order for the receivers query to even see the message at all.

The sender is able to upload their Dpush site and then disconnect from the network. They do not need to be online anymore until they wish to query the network to retrieve any messages that are waiting for them in the DHT.

**5. Dmail**

Dmail is yet again just a simple additional layer. Now on top of Dpush, which itself is on top of TargetedBlock. Dmail is a full e-mail replacement. It is not just decentralized spam resistant unsolicited messaging, but decentralized spam resistant automatically encrypted and inherently authenticated messaging. Secure enough for Edward Snowden, yet simple and more importantly fool-proof enough for a child. This completely deprecates e-mail as well as e-mail + GPG. A robust implementation with a fairly polished interface is already implemented in MORPHiS today.

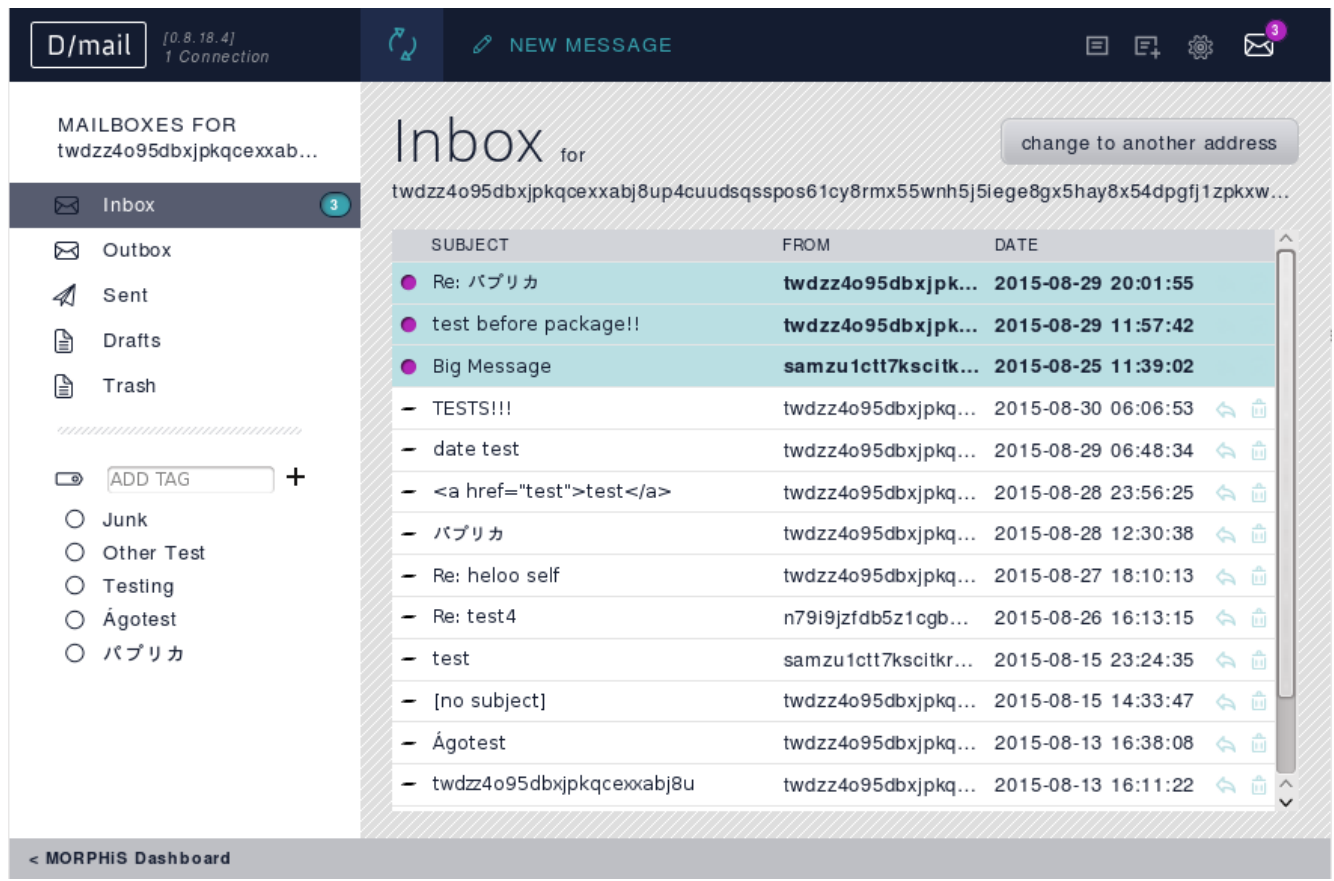

Dmail works by utilizing Dpush's updateable key published address meta-data (Dpush site / Dpush address) concept described in Section 4, and extends the information stored in the JSON meta-data to include additional information published by the address owner. In addition to the target_key and difficulty settings, Dmail also stores the name of the expected encryption mechanisms, as well an ephemeral Diffe-Hellman key used to generate a shared secret used by the symmetric encryption (AES-

256 as currently implemented) that Dmail automatically always uses.

Authentication is automatic and inherent with the very addresses used in Dmail. The sender field of a Dmail message is simply the sender's public key. The destination field is a hash of the receivers public key. The encrypted Dmail data stored in the TargetedBlock contains a signature of the Dmail data using the senders private key. For each user, their public key component is hashed, and that hash value is their Dmail address (Dpush address) itself. In this way, when a message says that it is from a certain Dmail address, one can inherently be sure that the owner (or controller) of that Dmail address sent the message, because the signature of the data must be by a key that has that address as the hash of the public component of the asymmetric key that signed the data. Thus, the keys used for authentication are inherent in the addresses in the Dmail protocol themselves. The two participants only need to know each others Dmail address, and are provided with full authentication as well as automatic and transparent encryption. No separate keys are needed to be exchanged for authentication, and the encryption keys are automatically generated per Dmail sent. As well, since the network hosts the Dmail address and site, the address meta-data, the receiver does not have to be online at the time a message is sent.

The protocol is a simple one way operation, with no back and forth between parties being needed. This is enabled by the fact that the DHT itself is able to act as the intermediary in a decentralized manner. The receiver originally has to have uploaded their Dmail site to the network. This is automatic upon address creation. After that point they can disconnect from the network. The sender then fetches the Dmail site from the network by using the receivers Dmail address as the ID they query the DHT network with. The network responds with the address meta-data (Dmail site) signed by the receiver. The sender is then able to validate that data against the very Dmail address value that they have for that receiver, as the data is signed, and the response includes the public key that signed it, and the very address queried is a hash of that public key. The sender then looks into the data to see the required target_key value, difficulty, as well as any encryption information needed. The sender then creates a Dmail data which is their message and includes their public key and the destination address. They then sign that data with their public key, and then they encrypt that data with a shared secret calculated by completing their half of the DH protocol with the ephemeral key that was specified in the receivers Dmail site. They then place that encrypted data into a DmailWrapper structure along with their public DH number. They then attach the DmailWrapper to a TargetedBlock. They then generate a valid header for the desired receiver's target_key for the TargetedBlock as the proof of work operation. At that point the sender simply issues a DHT put operation with the TargetedBlock as the data. If the TargetedBlock is valid and includes enough proof of work for the hash of its header to be within difficulty distance of the receivers target_key, then the block is inherently stored within the key space that the receiver will query. The receiver need not be online at the time, as it is the network that stores the TargetedBlock, as well, the network was where the receivers Dmail site is stored. Later when the receiver wants, they connect to the network and query their target_key key space with a ranged DHT get operation.

**6. MORPHiS**

The scope of MORPHiS is beyond what is reasonable to fully include in this paper. However, a brief description will follow in this section.

MORPHiS is an already published, functioning and deployed network. It is a very high performance Sybil[3] resistant custom and enhanced Kademlia[1] based DHT.

MORPHiS is passive monitoring proof, as it is fully encrypted using an optimized (light-weight channels and no rwindow) SSHv2 implementation as its node protocol, and with the Kademlia routing metric as who it will route to, combined with that the Node ID is a hash of its RSA-4096 public key used to authenticate its SSH connections. The Kademlia implementation in MORPHiS diverges greatly from the Kademlia paper and most if not all implementations of Kademlia. This is because the MORPHiS implementation is TCP connection based, with tunneling, instead of the connectionless recursive original design of Kademlia. MORPHiS also uses the hash of the node's RSA-4096 public key as its node ID, making it difficult to choose an arbitrary node ID, thus increasing the monitoring resistance as well as Sybil resistance. All hashes are done with the SHA-512 algorithm.

MORPHiS is at least weakly anonymous, as it tunnels through its directly connected peers, with the responders deeper in the tunnel only seeing the request as coming from their directly connected peers. This tunneling provides some obfuscation, as well as deniability about requests. However, as with other such designs, it is vulnerable to statistical correlation attacks. The important point however is that MORPHiS takes no extra percussion against such correlation attacks, or de-anonymizing attacks in general. This is because it is purposely designed and built to be as high performance of a base layer as is reasonably possible. The only compromise to its performance that it makes is the anti-entrapment feature, which is that a node encrypts the data blocks it stores and throws out the key so that it is only storing random bits it cannot decrypt. The requester of a data decrypts the data upon receiving it. This has only the O(1) cost of the extra layer of AES-256 encryption; it does not affect the efficiency or scaling of the network. The logical vision is that an anonymous layer can be built on top of such a high performance network as is MORPHiS, just as the attempted anonymous layers that are Tor, Freenet, and I2P are built upon the completely non-anonymous network that is the internet. MORPHiS, however, is designed from the beginning to be 100% Tor compatible, and is designed to not leak any information that would compromise the users anonymity under such use. It already works well over Tor by wrapping with proxychains, and also torsocks. Torsocks itself currently has a bug that it crashes unless you disable the MORPHiS HTTP interface (Maalstroom) with the --dm parameter. SOCKS5 support will be added shortly so that such a wrapper is no longer needed.

MORPHiS implements static keys (request ID is a hash of the data), updateable keys (request ID is a hash of the public component of the asymmetric key that signed the data), prefix searches (need only specify the prefix of a ID in order for the network to find it), and the TargetedBlock feature described in Section 3. The key space in all cases is an SHA-512 hash.

MORPHiS includes an HTTP interface, a text SSH console interface, as well as the binary node SSH interface. The HTTP interface is called Maalstroom. It is the most complete interface for end users at the moment. It provides an upload and download interface, an interface for general MORPHiS web browsing, as well as a robust and polished Dmail implementation and interface that will be familiar to any web mail user. It is ready for use today. In the future, a POP3 or even IMAP interface can be added. Using the Maalstroom mail interface, E-mail is now already deprecated; E-mail + GPG as well.

MORPHiS is programmed in 100% Python. It uses Python asyncio + co-routines in order to perform extremely well across platforms and out of the control of NSA and other state agency owned corporations (Prism + National Security Letter) like Google (Go) and Oracle (Java). Original plans were to port it to Rust-lang when Rust is ready; however, the Python asyncio implementation performed far beyond expectations, so such a port is now on the back burner. MORPHiS is licensed under the GPLv2, and available free to everyone.

MORPHiS is available at [https://morph.is](https://morph.is); as well as from within the MORPHiS network itself at the sp1nara3xhndtgswh7fznt414we4mi3y updateable key controlled by me.

**7. Conclusion**

In this paper has been presented the novel invention that is Dpush: A scalable decentralized spam resistant unsolicited messaging protocol. Also described is a simple layer on top called Dmail that deprecates e-mail as well as e-mail + GPG by providing fool-proof automatic encryption and inherent authentication on top of Dpush. Also pointed to is a mature implementation of Dmail on top of a Sybil resistant and as anonymous as Tor if used over Tor DHT implementation implementing the necessary features (TargetedBlock and prefix search) to support Dpush and thus Dmail. This is the start of the necessary work in order to help restore true democracy, and free the human species from the parasitical entities wielding information asymmetry against us in order to enslave us for their own inherently corrupt and degenerate will.